\documentclass[aps,showpacs,preprintnumbers,amsmath,amssymb]{revtex4}
\usepackage{dcolumn}
\usepackage{graphicx}

\usepackage[dvips]{epsfig}
\usepackage{amssymb}
\usepackage{color}
\usepackage{enumerate}
\usepackage{subfigure}
\usepackage{multirow}
\usepackage{diagbox}


\begin{document}
\baselineskip=0.8 cm
\title{\bf Magnetic reconnection and energy extraction from a Konoplya Zhidenko rotating non Kerr black hole}

\author{Fen Long$^{1}\footnote{Corresponding author: lf@usc.edu.cn}$,
Shangyun Wang$^{2}\footnote{sywang@hynu.edu.cn}$,
Songbai Chen$^{3,4}$\footnote{csb3752@hunnu.edu.cn},
Jiliang Jing$^{3,4}$ \footnote{jljing@hunnu.edu.cn}}
\affiliation{ $ ^1$ School of Mathematics and Physics, University of South China,
Hengyang, 421001, People's Republic of China
\\$ ^2$ College of Physics and Electronic Engineering, Hengyang Normal University, Hengyang 421002, People's Republic of China
\\$ ^3$ Department of Physics, Institute of Interdisciplinary Studies, \
\\ Key Laboratory of Low Dimensional Quantum Structures and Quantum Control of Ministry of Education,
\\Synergetic Innovation Center for Quantum Effects and Applications, Hunan
Normal University,  Changsha, Hunan 410081, People's Republic of China
\\
$ ^4$Center for Gravitation and Cosmology, College of Physical Science and Technology, Yangzhou University, Yangzhou 225009, People's Republic of China}

\begin{abstract}
\baselineskip=0.6 cm
\begin{center}
{\bf Abstract}
\end{center}

Recently, magnetic reconnection has attracted considerable attention as a novel energy extraction mechanism, relying on the rapid reconnection of magnetic field lines within the ergosphere. We have investigated the properties of the energy extraction via magnetic reconnection in a Konoplya-Zhidenko rotating non-Kerr black hole spacetime with an extra deformation parameter.  Our results show that the positive deformation parameter expands the possible region of energy extraction and improves the maximum power, maximum efficiency, and the maximum ratio of energy extraction between magnetic reconnection and the Blandford-Znajek process.  This means that in the Konoplya-Zhidenko rotating non-Kerr black hole spacetime one can extract more energy via magnetic reconnection  than in the Kerr black hole case. These effects of the deformation parameter may provide valuable clues for future astronomical observations of black holes and verification of gravity theories.

\end{abstract}

 \maketitle
\newpage
General relativity theoretically predicts the existence of black holes, and observations of gravitational waves \cite{L1,L2,L3,L4,L5} as well as black hole images \cite{EHT1,EHT2} provide favorable experimental evidence for confirming their existence, which ushering in a golden age of black hole physics research. Black holes, as the most fascinating and intriguing entities in astrophysics, are notable not only for their extremely curved geometry and intensely strong gravitational field but also for their pivotal role in explaining high-energy astrophysical phenomena, such as active galactic nuclei (AGN)\cite{AGN1,AGN2,AGN3,AGN4}, gamma-ray bursts (GRB) \cite{GRB1,GRB2,GRB3}, and ultraluminous X-ray binaries \cite{Xray}, where an enormous amount of energy is released. Exploring the mechanisms of energy extraction from black holes helps to uncover the underlying principles behind the generation of high-energy astrophysical phenomena.

The most famous mechanism for extracting energy from black holes is the Penrose process \cite{pp}, where the energy is extracted from a rotating black hole through particle splitting. The incoming particle enters the ergosphere and splits into two fragments. One fragment with negative energy crosses the outer horizon and falls into the black hole, whereas the other escapes to infinity carrying more energy than the original particle.
However, this process requires that the relative three-velocity between the particles after split must satisfy the condition $v> c/2$ \cite{pp1}, which is relatively difficult to achieve experimentally. At the same time, in order to achieve the highest possible energy extraction efficiency, particles split as close to the event horizon as possible. Since these conditions are difficult to achieve and sustain in nature, the Penrose process is insufficient to explain high-energy astrophysical phenomena. The magnetic Penrose process (MPP) \cite{Mpp} indicates that when a black hole is surrounded by a magnetic field, the relative velocity condition for particle splitting is easily satisfied, and the energy extraction efficiency of the MPP exceeds $100\%$ in realistic settings \cite{Mpp1,Mpp2,Mpp3}.
The Blandford-Znajek (BZ) \cite{BZ,BZ1} process utilizes the interaction between a black hole and its surrounding electromagnetic field to extract rotational energy from the rotating black hole, rather than through particle splitting. Therefore, the MPP and BZ process are considered as excellent models for explaining the energy mechanism of active galactic nucleus jets and gamma-ray burst central engines. Furthermore, both the MPP and the BZ process depend on the assumption that the black hole is immersed in an external magnetic field.\cite{Mpp3,BZ1}.
In addition to the three mechanisms mentioned above, there are several well-known energy extraction mechanisms, such as collisional Penrose process \cite{CPP}, the magnetohydrodynamic Penrose process \cite{MHDPP}, Superradiant scattering \cite{SS} and so on.

Besides the previously mentioned mechanisms, recent researches (Koide and Arai \cite{KA}; Comisso and Asenjo \cite{CA}) proposed a new mechanism for energy extraction from the ergosphere of the rotating black holes via magnetic reconnection. Magnetic reconnection occurs in the plasma rotating around a black hole in the equatorial plane.
The frame-dragging effect generated by rapidly spinning black hole creates adjacent antiparallel magnetic field on the equatorial plane. This creates an equatorial current sheet is prone to fast plasmoid-mediated magnetic reconnection, which converts magnetic energy into kinetic energy of the plasma and generates a pair of plasma outflows in opposite directions in the reconnection layer \cite{CA}. One part of the plasma carries a lot of negative energy and falls into the black hole event horizon along retrograde orbit, while the other part of the plasma with positive energy escapes to infinity along prograde orbit, thus achieving energy extraction.
Energy extraction through magnetic reconnection relies on local magnetic fields and antiparallel magnetic field structures, without the need for the large-scale magnetic field backgrounds required in the MPP and the BZ process, making it more likely to occur in the presence of a significant magnetic field ring component in the black hole ergosphere. Comisso and Asenjo show that in Kerr black holes with high plasma spin and strong magnetization, the maximum efficiency of energy extraction through magnetic reconnection can reach $3/2$ \cite{CA}. Its energy extraction power can also exceed that of the BZ process in certain cases, making it a promising method for black hole energy extraction\cite{CA}. Researchs on energy extraction via magnetic reconnection have extended to various rotating compact objects \cite{1,2,3,4,5,6,7,8,9,10,11,12,13,14,15}, which provides valuable material for verifying gravity theories.

In this paper, we will study energy extraction via magnetic reconnection in Konoplya-Zhidenko rotating non-Kerr black hole \cite{KZ1}.
The Konoplya-Zhidenko rotating non-Kerr metric exhibits some non-negligible deviation from Kerr spacetime. It has been found that the Kerr metric with these deviations can likewise produce the same frequencies of black-hole ringing \cite{KZ1}, which preserves an opportunity for exploring alternative theories beyond general relativity.
Additionally, the study of quasi periodic oscillation constraints \cite{KZ2} and the iron lines \cite{KZ3} further suggests that real astrophysical black holes can be described by the Konoplya-Zhidenko rotating non-Kerr metric.
Therefore, it is necessary to further explore the features of this rotating non-Kerr metric through various processes to distinguish this theory from general relativity. The energy extraction via the Penrose process in Konoplya-Zhidenko rotating non-Kerr black holes \cite{KZ4} show that he maximum efficiency of the energy extraction process of black hole with superspinning is greatly improved when the deformation parameter approaches zero, which is much higher than the Kerr case.
Therefore, we will investigate energy extraction via magnetic reconnection in the Konoplya-Zhidenko rotating non-Kerr black hole, analyzing the impact of the deformation parameter on energy extraction efficiency. We will then compare the results with those from energy extraction via magnetic reconnection in the Kerr black hole and the Penrose process in a Konoplya-Zhidenko rotating non-Kerr black hole. This comparison will allow us to examine the differences in energy extraction between these different black holes and processes, and to provide insights into their distinct characteristics, which offers additional information and multiple perspectives to test the deviation from Kerr black hole in general relativity.

The paper is organized as follows. In Sec. II, we briefly review the metric of the Konoplya-Zhidenko rotating non-Kerr spacetime and calculated several characteristic orbital radii related to energy extraction, including the horizon radius, the redshift radius, and the photosphere radius. In Sec.III, we investigate the magnetic reconnection of Konolya-Zhidenko rotating non Kerr black hole and clarify the possible parameter regions where energy extraction can occur. In Sec.IV, we calculate the power and efficiency via magnetic reconnection. Finally, we present a brief summary.

\section{The Konoplya-Zhidenko rotating non-Kerr black hole spacetime}

The metric of a Konoplya-Zhidenko rotating non-Kerr black hole can be expressed in the Boyer-Lindquist coordinates as follows\cite{KZ1}:
\begin{eqnarray}
\begin{aligned}
d s^2 & =-\left(1-\frac{2 M r+\frac{\eta}{r}}{\rho^2}\right) d t^2+\frac{\rho^2}{\Delta} d r^2+\rho^2 d \theta^2+\frac{A}{\rho^2}\sin ^2 \theta d \phi^2 -\frac{2\left(2M r+\frac{\eta}{r}\right) a \sin ^2 \theta}{\rho^2} d t d \phi,
\end{aligned}\label{metric0}
\end{eqnarray}
with
\begin{eqnarray}
\begin{aligned}
\rho^2=r^2+a^2 \cos ^2 \theta, \quad \Delta=r^2-2 M r+a^2-\frac{\eta}{r}, \quad A=(r^2+a^2)^2-a^2\Delta \sin^2\theta,
\end{aligned}
\end{eqnarray}
where $M$ and $a$ represent the mass and the rotation parameter of black hole, respectively. The deformation parameter $\eta$, which characterizes deviations from Kerr spacetime, causes the metric to reduce to the Kerr case when $\eta$ is absent. The presence of the deformation parameter $\eta$ will further influence the magnetic reconnection process by modifying the spacetime properties in the strong field region.

Initially, we will perform a detailed analysis of a series of characteristic surfaces and their orbital radii related to magnetic reconnection and energy extraction, including the horizon, the infinite redshift surface, and the photosphere. The region of the ergosphere defined by the outer horizon and the outer infinite redshift surface is where magnetic reconnection occurs, making it crucial for extracting rotational energy from a rotating black hole. For this rotating non-Kerr black hole, the outer horizon radius is given by the largest solution of the following equation
\begin{eqnarray}
r^3-2 M r^2+a^2 r-\eta& =& 0,
\end{eqnarray}
and the outer infinite redshift surface radius is given by the largest solution of the equation
\begin{eqnarray}
r^3-2 M r^2+a^2 r \cos^2{\theta}-\eta& =& 0.
\end{eqnarray}
The dependence of the outer horizon and the outer infinite redshift surface on the deformation parameter $\eta$ in this black hole was discussed in\cite{KZ4}. As $\eta$ increases, the ergosphere in the equatorial plane becomes thinner.
When $a>M$ and positive $\eta$ approaches zero, the outer horizon radius becomes very small, resulting in a much thicker ergosphere in the equatorial plane compared to the $a < M$ case, and the maximum efficiency of Penrose process in this black hole becomes almost unlimited.

Another key orbit in magnetic reconnection is the photon orbit with $\mu=0$.
The Hamiltonian for photon propagation along null geodesics can be written as
\begin{equation}
H(x,p)=\frac{1}{2}g^{\mu \nu}(x)p_\mu p_\nu = 0.
\end{equation}
There are two conserved quantities for photon propagation: $E$, the energy of the photon, and $L_{\phi}$, which are expressed as
\begin{eqnarray}
E &=& -p_{t}=-g_{tt} \dot{t}-g_{t\phi} \dot{\phi},\quad\quad \quad\nonumber \\
L_{\phi}&=& p_{\phi}=g_{t\phi}\dot{t}+g_{\phi\phi}\dot{\phi}.
\end{eqnarray}
Using these conserved quantities, the null geodesics in a Konoplya-Zhidenko rotating non-Kerr black hole spacetime can be written as:
\begin{eqnarray}
\frac{dt}{d\tau} & = &E+\frac{\left(a^2 E-a L_\phi+E r^2\right)\left(2 M r^2+\eta\right)}{\Delta \rho^2 r} \\
\frac{d\phi}{d\tau} & =&\frac{a E \sin ^2 \theta\left(2 M r^2+\eta\right)+a^2 L_\phi r \cos ^2 \theta-L_\phi\left(2 M r^2-r^3+\eta\right)}{\Delta \rho^2 r \sin ^2 \theta}, \\
\rho^4 \left(\frac{dr}{d\tau}\right)^2 & =&R(r)=-\Delta K+\left[a L_\phi-\left(r^2+a^2\right) E\right]^2, \\
\rho^4 \left(\frac{d\theta}{d\tau}\right)^2 & =&\Theta=K-\frac{1}{\sin^2\theta}(L_\phi-a E \sin^2\theta)^2,\label{theta}
\end{eqnarray}
where $K$ is a constant related to the Carter constant. According to the conditions for photon circular orbits $R(r)=0$ and $R'(r)=0$, combined with the definition of impact parameters $\xi=\frac{L_\phi}{E}$ and $\sigma=\frac{K}{E^2}$, one can obtain
\begin{equation}
\begin{aligned}
\xi & =\frac{a^2\left(\eta-2 M r^2-2 r^3\right)+6 M r^4+5 \eta r^2-2 r^5}{a\left(2 r^3-2 M r^2+\eta\right)}, \\
\sigma & =\frac{16 r^5\left(r^3+a^2 r-2 M r^2-\eta\right)}{\left(2 r^3-2 M r^2+\eta\right)^2}\label{constant}.
\end{aligned}
\end{equation}
For the case of photon propagates on the equatorial plane, its motion satisfies
\begin{eqnarray}
\Theta|_{\theta=\frac{1}{2}}=0.\label{equatorial}
\end{eqnarray}
From Eqs. (\ref{theta}), (\ref{constant}) and (\ref{equatorial}), we find that the radius $r_{ph}$ of the unstable photon circular orbits on the equatorial plane can be given by
\begin{eqnarray}
(2r^3-6Mr^2-5\eta)^2-8a^2(2Mr^3+3\eta r)=0.\label{rph}
\end{eqnarray}
This study is based on the simplified assumption that magnetic reconnection takes place in plasma rotating along circular orbits around a black hole in the equatorial plane.
According to the above conditions for circular orbits and the $r$-component of the Euler-Lagrange equation, the Keplerian angular velocity $\Omega_{K}=\frac{d\phi/d\tau}{dt/d\tau}$ can be expressed as
\begin{eqnarray}
\Omega_{K}=\frac{\sqrt{2Mr^2+3\eta}}{a\sqrt{2Mr^2+3\eta}\pm \sqrt{2r^{5}}}.\label{Omega}
\end{eqnarray}
The upper sign indicates prograde orbits, whereas the lower sign corresponds to retrograde orbits.

\section{ENERGY EXTRACTION VIA MAGNETIC RECONNECTION MECHANISM}
In this section, we will investigate the energy extraction of the Konoplya-Zhidenko rotating non-Kerr black hole via magnetic reconnection, and analyze the influence of the deformation parameter on energy extraction and further explore the differences by comparing the research results with those in the Kerr case. For convenience, the plasma energy density of magnetic reconnection will be calculated and analyzed in the Zero Angular Momentum Observer (ZAMO) frame.
The line element in this frame appears as in Minkowski spacetime, written as
\begin{equation}
d s^2=-d \hat{t}^2+\sum_{i=1}^3\left(d \hat{x}^i\right)^2=\eta_{\mu \nu} d \hat{x}^\mu d \hat{x}^\nu,
\end{equation}
The transformation between the Boyer-Lindquist(BL) frame$(dt,dx^1,dx^2,dx^3)$ and the ZAMO frame$(d\hat{t},d\hat{x}^1,\\d\hat{x}^2,d\hat{x}^3)$ is expressed as
\begin{equation}
d \hat{t}=\alpha d t, \quad d \hat{x}^i=\sqrt{g_{i i}} d x^i-\alpha \beta^i d t,\label{transformation}
\end{equation}
where
\begin{eqnarray}
\alpha  =\sqrt{-g_{t t}+\frac{g_{\phi t}^2}{g_{\phi \phi}}}=\left(\frac{\Delta \rho^2}{A}\right)^{1/2},\quad
\beta^\phi  =\frac{\sqrt{g_{\phi \phi}} \omega^\phi}{\alpha}=\frac{\omega^\phi}{\alpha}\left(\frac{A}{\rho^2}\right)^{1/2}\sin\theta,
\end{eqnarray}
here $\alpha$ is the lapse function, $\beta^i=\left(0,0, \beta^\phi\right)$ is the shift vector, and $\omega^\phi=-g_{\phi t}/g_{\phi \phi}$ means the angular velocity of the frame dragging. According to Equation (\ref{transformation}), we can obtain the transformation of vectors between the two frames as follows:
\begin{equation}
\hat{\psi}^0=\alpha \psi^0 ,\quad \hat{\psi}^i=\sqrt{g_{i i}} \psi^i-\alpha \beta^i \psi^0 .
\end{equation}

During magnetic reconnection, the breaking and reconnection of magnetic field lines release a substantial amount of energy, and the plasma escapes from the reconnection layer with this energy. When part of the plasma carrying negative energy fall into the black hole, while another part carrying positive energy manages to escape the black hole and travel to infinity, it is possible to steal energy from the black hole. Next, we would like to analyze the feasibility conditions for energy extraction from the Konoplya-Zhidenko rotating non-Kerr black hole via magnetic reconnection. Under the one-fluid approximation, the energy-momentum tensor $T^{\mu \nu}$ of the plasma can be written as
\begin{equation}
T^{\mu \nu}=p g^{\mu \nu}+w U^\mu U^\nu+F_\delta^\mu F^{\nu \delta}-\frac{1}{4} g^{\mu \nu} F^{\rho \delta} F_{\rho \delta},
\end{equation}
here, $p$, $w$, $U^\mu$ and $F^{\mu \nu}$ respectively represent plasma pressure, enthalpy density, 4-velocity and electromagnetic tensor.
According to definition of the ``energy-at-infinity" density, we have
\begin{equation}
e^{\infty}=-\alpha g_{\mu 0}T^{\mu 0}=\alpha \hat{e}+\alpha \beta^\phi \hat{P}^\phi,
\end{equation}
where the total energy density $\hat{e}$ and the azimuthal component of the momentum density $\hat{P}^\phi$ are given by
\begin{equation}
\begin{aligned}
\hat{e} & =w \hat{\gamma}^2-p+\frac{\hat{B}^2+\hat{E}^2}{2}, \\
\hat{P}^\phi & =w \hat{\gamma}^2 \hat{v}^\phi+(\hat{B} \times \hat{E})^\phi,
\end{aligned}
\end{equation}
here, $\hat{\gamma}=\hat{U}^0=\left[1-\sum_{i=1}^3\left(d \hat{v}^i\right)^2\right]^{-1 / 2}$, $\hat{B}^i=\epsilon^{i j k} \hat{F}_{j k} / 2 $, $\hat{E}^i=\eta^{i j} \hat{F}_{j 0}=\hat{F}_{i 0}$ and $\hat{v}^\phi$ respectively represent the Lorentz factor, the components of the electric and magnetic fields, and the velocity of the plasma flowing out of the reconnection layer in the ZAMO frame.

The energy-at-infinity density could be separated into $e^{\infty}=e_{hyd}^{\infty}+e_{em}^{\infty}$, where $e_{hyd}^{\infty}$ and $e_{em}^{\infty}$ represent the hydrodynamic component and the electromagnetic component, respectively, and they are given by
\begin{eqnarray}
e_{h y d}^{\infty}&=&\alpha \hat{e}_{h y d}+\alpha \beta^\phi \omega \hat{\gamma}^2 \hat{v}^\phi, \\
e_{e m}^{\infty}&=&\alpha \hat{e}_{em}+\alpha \beta^\phi(\hat{B} \times \hat{E})_\phi.
\end{eqnarray}
$\hat{e}_{h y d}$ and $\hat{e}_{em}$ are respectively represent the hydrodynamic and electromagnetic energy densities in the ZAMO frame. Considering that in the magnetic reconnection process, most of the magnetic energy is converted into the kinetic energy of the plasma, then the contribution of $e_{em}^{\infty}$ in the total energy can be ignored, which leads to $e^{\infty}\approx e_{hyd}^{\infty}$. Additionally, assuming that the plasma element is incompressible and adiabatic, the energy-at-infinity density can be rewritten as\cite{KA}
\begin{equation}
e^{\infty}=\alpha \omega \hat{\gamma}\left(1+\beta^\phi \hat{v}^\phi\right)-\frac{\alpha p}{\hat{\gamma}} .\label{einfty}
\end{equation}
For the convenience of analyzing local reconnection process, one can introduce the local rest frame $x^{\mu}{'} = (x^{0}{'},x^{1}{'}, x^{2}{'},x^{3}{'})$ to study the bulk plasma rotating with Keplerian angular velocity in the equatorial plane, where the directions of $x^{1}{'}$ and $x^{3}{'}$ in the local rest frame are parallel to the radial $x^{1}= r$ and azimuthal direction $x^{3}= \phi$, respectively. According to the transformation Equation (\ref{transformation}), the Keplerian angular velocity observed in the ZAMO frame is given by
\begin{equation}
\hat{v}_K=\frac{d \hat{x}^\phi}{d \hat{x}^t}=\frac{\sqrt{g_{\phi \phi}} d x^\phi-\alpha \beta^\phi d x^t}{\alpha d x^t}=\frac{\sqrt{g_{\phi \phi}}}{\alpha} \Omega_K-\beta^\phi .
\end{equation}

Considering "relativistic adiabatic incomprehensible ball approach" and assuming a relativistically hot plasma with a polytropic index $\Gamma=3/4$, the energy at infinity per enthalpy can be expressed in accordance with $\epsilon_{ \pm}^{\infty}=e^{\infty}/w$ as follows\cite{CA}:
\begin{equation}
\epsilon_{ \pm}^{\infty}=\alpha \hat{\gamma}_K\left[\left(1+\beta^\phi \hat{v}_K\right) \sqrt{1+\sigma_0} \pm \cos \xi\left(\beta^\phi+\hat{v}_K\right) \sqrt{\sigma_0}-\frac{\sqrt{1+\sigma_0} \mp \cos \xi \hat{v}_K \sqrt{\sigma_0}}{4 \hat{\gamma}_K^2\left(1+\sigma_0-\cos ^2 \xi \hat{v}_K^2 \sigma_0\right)}\right],\label{epsilon}
\end{equation}
where symbols $+$ and $-$ represent plasma particles undergoing acceleration and deceleration, respectively escaping from the reconnection layer with co-rotating and counter-rotating outflow directions. The corresponding Lorentz factor denoted as $\hat{\gamma}_K=1/\sqrt{1-\hat{v}_{K}^{2}}$. The plasma magnetization upstream of the reconnection layer is given by $\sigma_0 = B_{0}^{2}/w_0$. From Eq.(\ref{epsilon}), we can deduce that the hydrodynamic energy at infinity per enthalpy is a function of the critical parameters ($a, M, \eta, \sigma_0, \xi, r$), where $r$ indicates the location of the reconnection point, denoted as the X-point. Additionally, $\xi$ represents the orientation angle between the magnetic field lines and the azimuthal direction in the equatorial plane. Many studies have shown that the energy-at-infinity per enthalpy via magnetic reconnection favors smaller values of the angle $\xi$ \cite{1,6,7}. Therefore, we sets $\xi=\pi/12$ to study the effect of deformation parameters $\eta$ on energy extraction. The existence of deformation parameters in the Konoplya-Zhidenko rotating non-Kerr black hole expands the range of the rotation parameter $a$, allowing it to be greater than $M$. When the rotation parameter $a>M$, the deformation parameter must satisfy the condition $\eta>0$; When the rotation parameter $a<M$, the deformation parameter must satisfy the condition $\eta>\eta_{c}=-\frac{2}{27}\left(\sqrt{4M^2-3a^2}+2M\right)^2\left(\sqrt{4M^2-3a^2}-M\right)$ to ensure the existence of the horizon and the ergosphere, which is crucial for energy extraction.

\begin{figure}[t]
\begin{center}
\includegraphics[width=5cm]{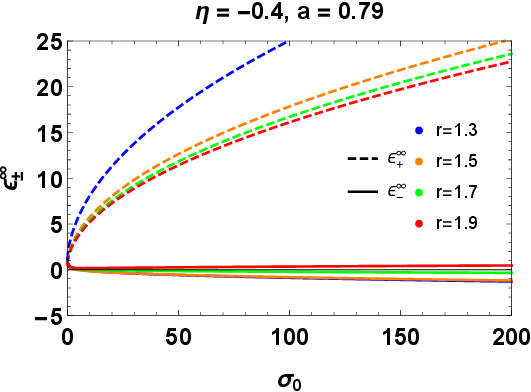}\includegraphics[width=5cm]{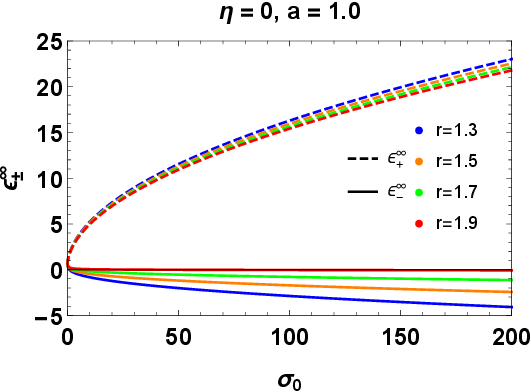}\includegraphics[width=5cm]{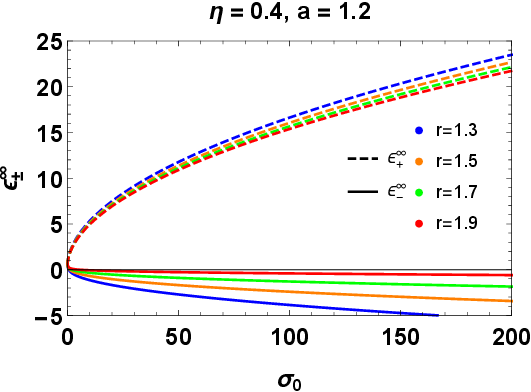}
\caption{ The change of the energy at infinity per enthalpy $\epsilon_{ \pm}^{\infty}$ with the plasma magnetization parameter $\sigma_0$ for different rotation parameter $a$, dominant X-point location $r$ and deformation parameter $\eta$, we set $M=1$.}\label{Fig1}
\end{center}
\end{figure}

In Fig.\ref{Fig1}, we present the change of the energy at infinity per enthalpy $\epsilon_{ \pm}^{\infty}$ with the plasma magnetization parameter $\sigma_0$ for different rotation parameter $a$, dominant X-point location $r$ and deformation parameter $\eta$. We take the value of $a$ and $\eta$ in the range where  the existence of the horizon is ensured. Fig.\ref{Fig1} shows that regardless of the values of the rotation parameter and the deformation parameter, the energy at infinity per enthalpy $\epsilon_{ +}^{\infty}$ increases with the plasma magnetization parameter $\sigma_0$, while $\epsilon_{ -}^{\infty}$ decreases except the region where $\sigma_0$ is small. Therefore, we assume that the ion magnetization parameter $\sigma_0=100$ to investigate the influence of deformation parameters $\eta$ on the energy at infinity per enthalpy $\epsilon_{ \pm}^{\infty}$.
\begin{figure}[ht]
\begin{center}
\subfigure[]{\includegraphics[width=6cm]{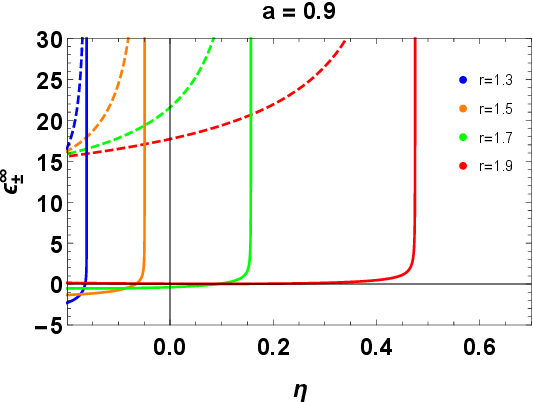}}\subfigure[]{ \includegraphics[width=6cm]{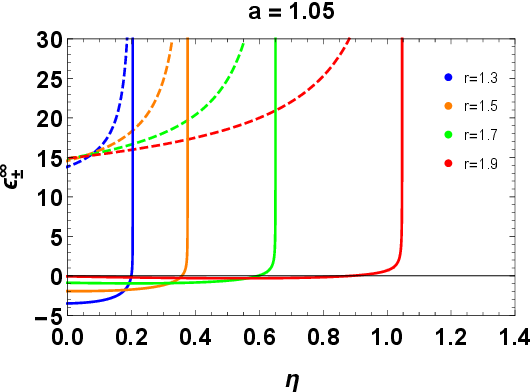}}\\

\subfigure[]{\includegraphics[width=6cm]{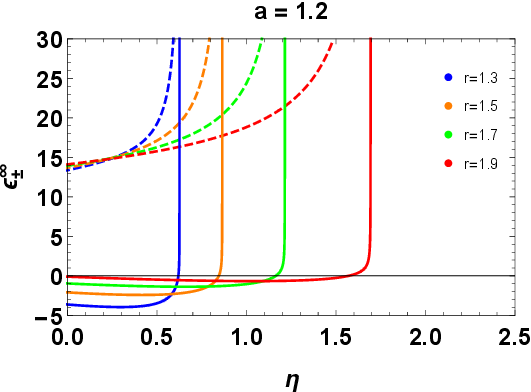}}\subfigure[]{\includegraphics[width=6cm]{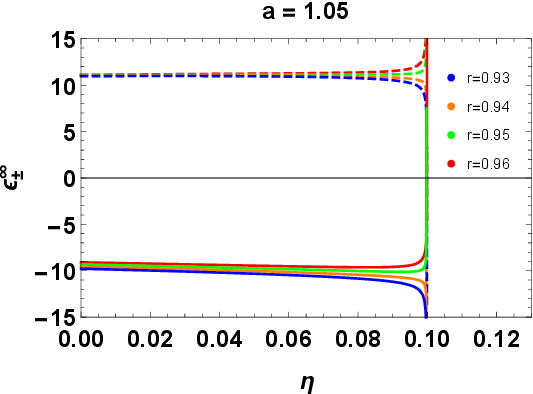}}
\caption{ The change of the energy at infinity per enthalpy $\epsilon_{ \pm}^{\infty}$ with the deformation parameter $\eta$ for different rotation parameter $a$ and dominant X-point location $r$. The dashed and solid lines correspond to the energy at infinity per enthalpy $\epsilon_{+}^{\infty}$ and $\epsilon_{-}^{\infty}$. We set $\sigma_0=100$ and $M=1$. }\label{Fig2}
\end{center}
\end{figure}

In Fig.\ref{Fig2}, we present the change of the energy at infinity per enthalpy $\epsilon_{ \pm}^{\infty}$ with the deformation parameter $\eta$ for different rotation parameter $a$. We found that when the value of the dominant X-point location $r$ is slightly greater than $M$, the energy at infinity per enthalpy of the accelerated plasma $\epsilon_{+}^{\infty}$ increases with the deformation parameters $\eta$;
the decelerated part $\epsilon_{-}^{\infty}$ slowly increases with the deformation parameter in the cases $a=0.9$ and $a=1.05$, but it first slowly decreases and then sharply increases with the deformation parameter in the case with $a=1.2$.
It is worth noting that when the rotation parameter $M<a\leq\frac{2\sqrt{3}}{3}M\approx1.155M$ and the dominant X-point location is slightly greater(or smaller) than $r_c$ which is determined by the maximum(or second largest) value in the numerical solution of equation $a\sqrt{2M r_c^3+3\eta_c r_c}+\sqrt{2}(r_c^3-2M r_c^2-\eta_c)=0$, the values of the energy at infinity per enthalpy of the accelerated plasma $\epsilon_{+}^{\infty}$ and the decelerated part $\epsilon_{-}^{\infty}$ are close to opposite numbers, so their sum tends to be close to zero. This implies that $\epsilon_{+}^{\infty}$ is much greater than the sum of $\epsilon_{+}^{\infty}$ and $\epsilon_{-}^{\infty}$, which is different from the characteristics in Kerr black holes and may introduce new features to the energy extraction results of the Konoplya-Zhidenko rotating non-Kerr black hole.
Similarly to the Penrose process, in order to extract energy from the black hole through magnetic reconnection, it is anticipated that the decelerated plasma should exhibit negative energy as observed at infinity, while the accelerated plasma is expected to have positive energy, surpassing its rest mass and thermal energy. Thus, the conditions of extracting energy can be expressed as \cite{CA}
\begin{equation}
\epsilon_{-}^{\infty}<0, \quad \text { and } \quad \Delta \epsilon_{+}^{\infty}=\epsilon_{+}^{\infty}-\left(1-\frac{\Gamma}{\Gamma-1} \frac{p}{\omega}\right)=\epsilon_{+}^{\infty}>0.\label{condition}
\end{equation}
From Fig.\ref{Fig2}, it can be seen that the value of the energy at infinity per enthalpy of the accelerated plasma $\epsilon_{+}^{\infty}$ in the Konoplya-Zhidenko rotating non-Kerr black hole is always positive, so we only need to consider the condition $\epsilon_{-}^{\infty}<0$.

\begin{figure}[ht]
\begin{center}
\includegraphics[width=4cm]{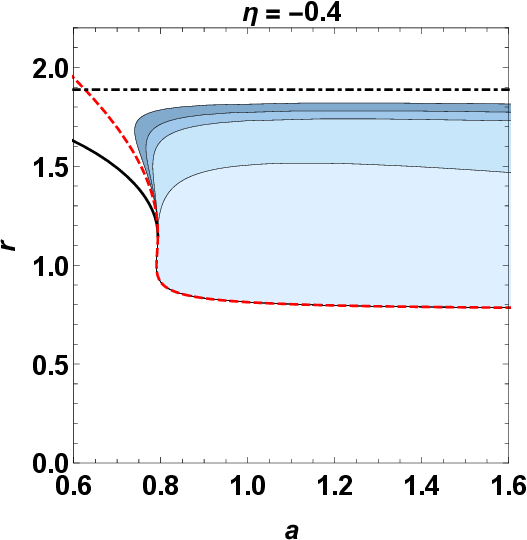}\includegraphics[width=4cm]{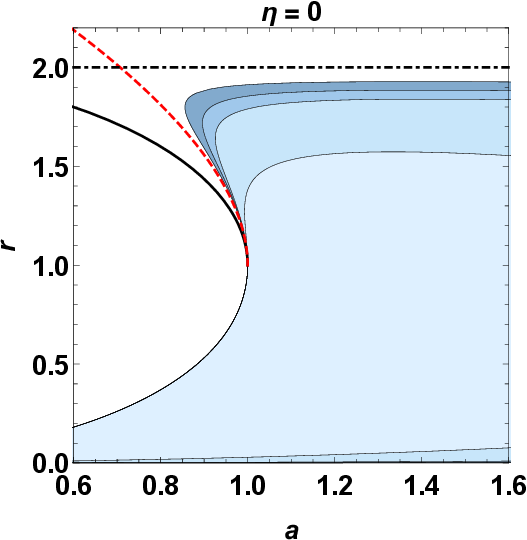}
\includegraphics[width=4cm]{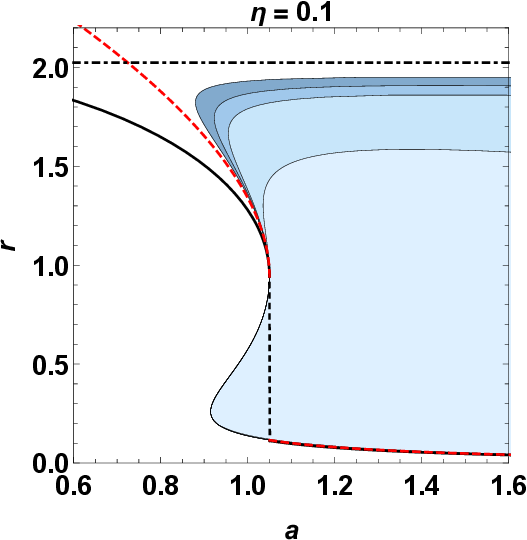}\includegraphics[width=4cm]{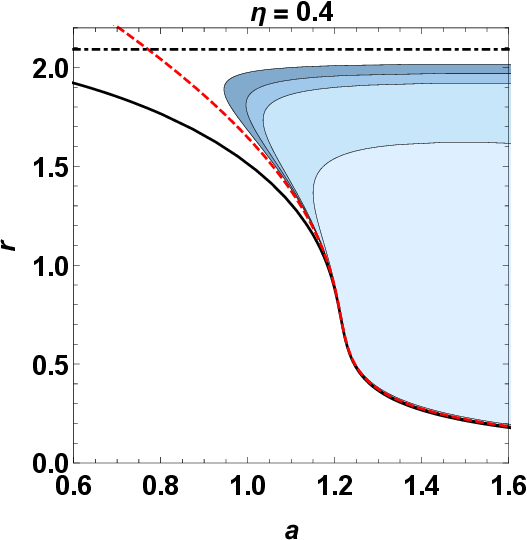}\\
\includegraphics[width=4cm]{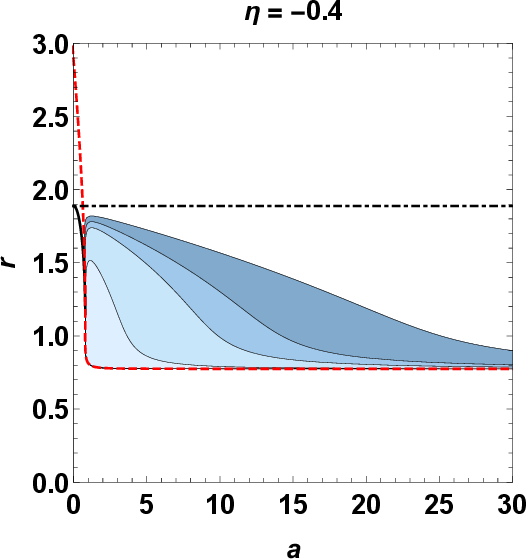}\includegraphics[width=4cm]{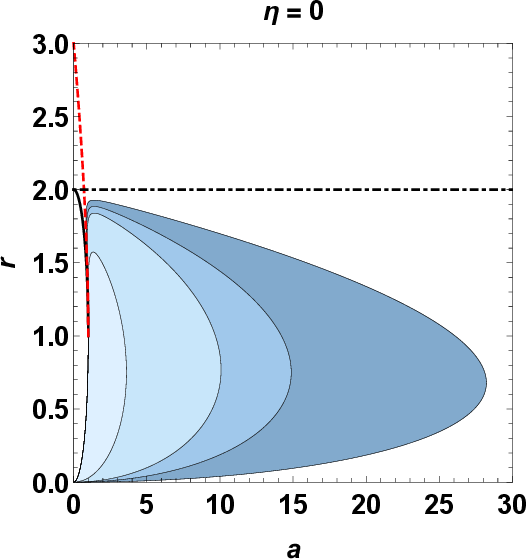}
\includegraphics[width=4cm]{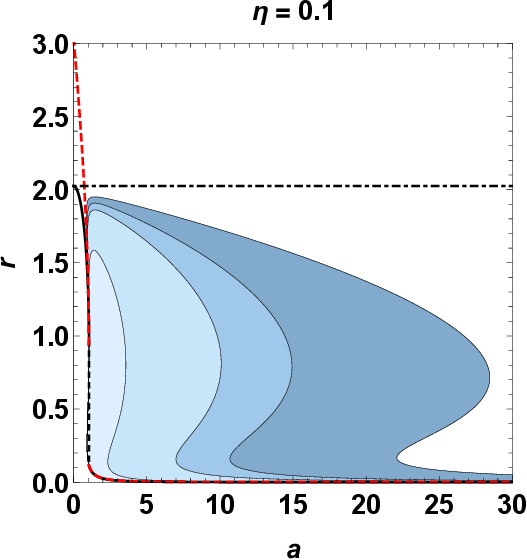}\includegraphics[width=4cm]{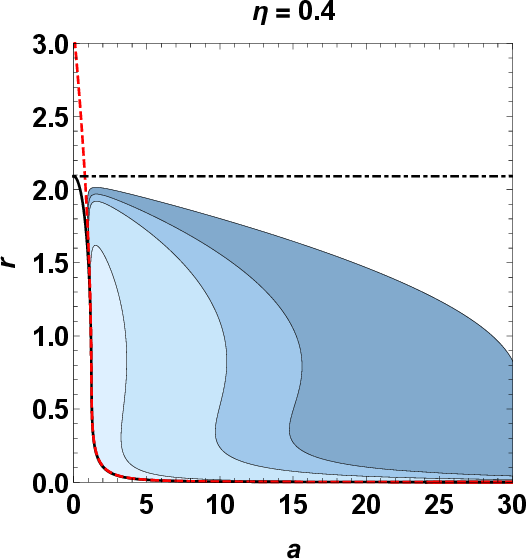}
\caption{ Regions of the phase space $(a,r)$ that satisfy the condition $\epsilon_{-}^{\infty}<0$. The blue shaded areas from light to dark correspond to $\epsilon_{+}^{\infty}>0$ for the plasma magnetization parameter $\sigma_0=1, 5, 10,100$. The thick black solid line, red dashed line, and dot-dashed line correspond to the radius of the outer horizon, photosphere, and the outer infinite redshift surface, respectively. We set $M=1$. }\label{Fig3}
\end{center}
\end{figure}

\begin{figure}[ht]
\begin{center}
\includegraphics[width=4cm]{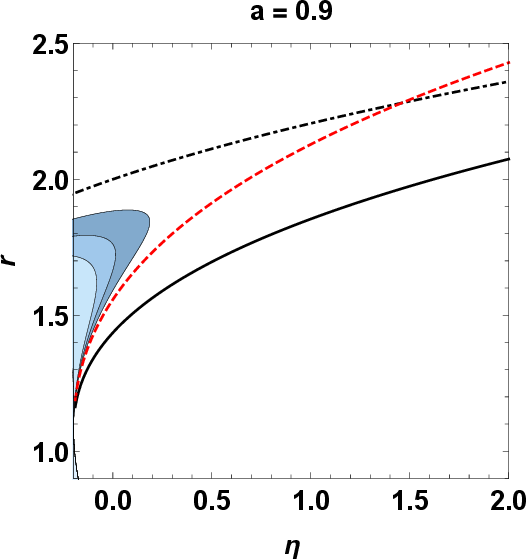}\includegraphics[width=4cm]{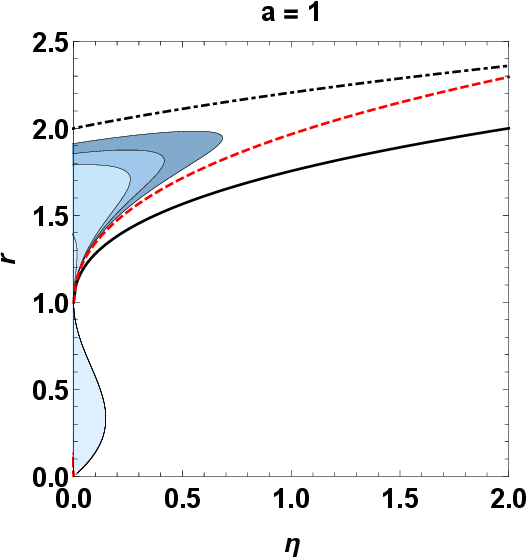}
\includegraphics[width=4cm]{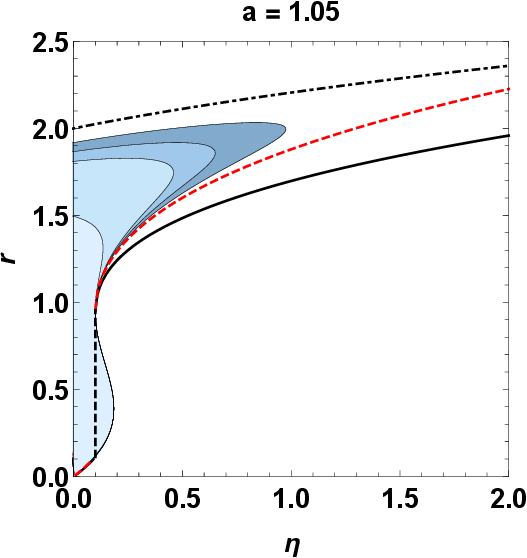}\includegraphics[width=4cm]{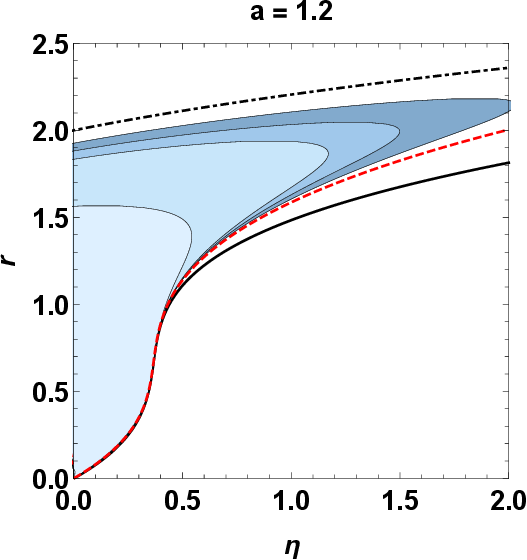}\
\includegraphics[width=4cm]{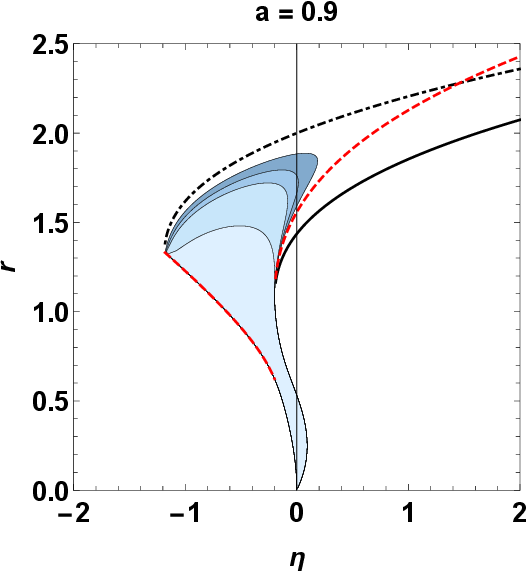}\includegraphics[width=4cm]{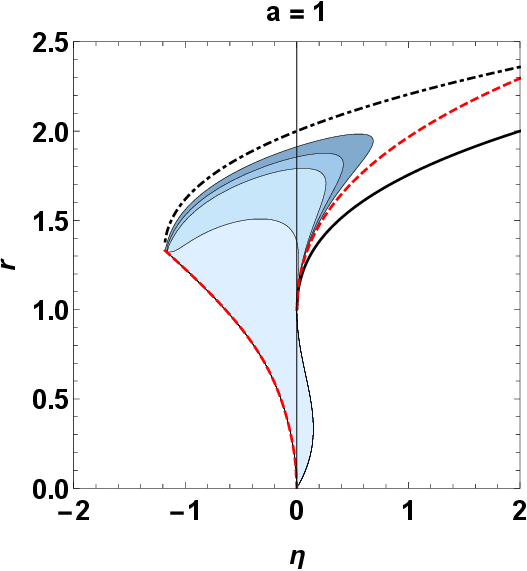}
\includegraphics[width=4cm]{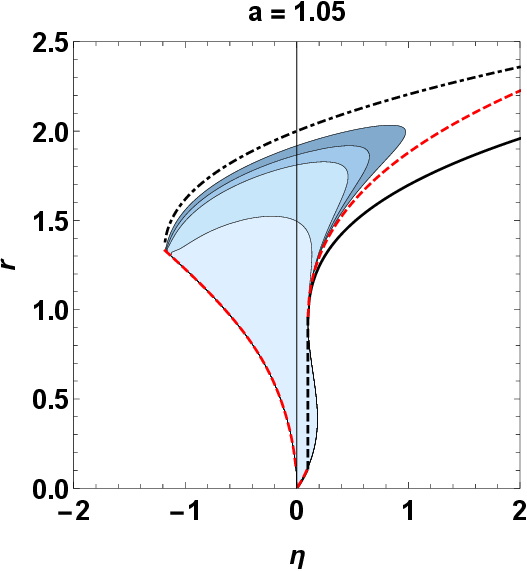}\includegraphics[width=4cm]{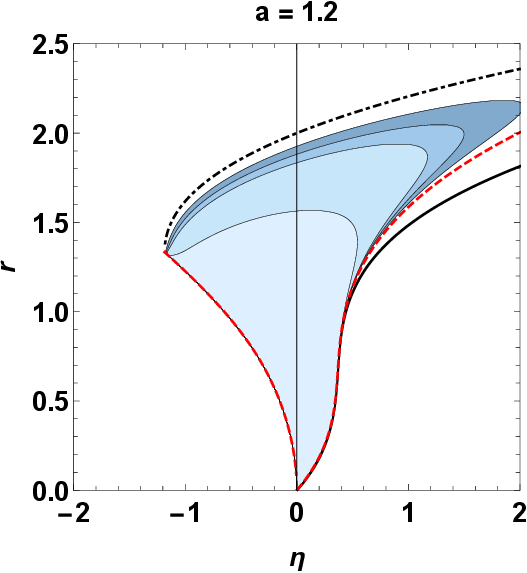}\

\caption{ Regions of the phase space $(\eta,r)$ that satisfy the condition $\epsilon_{-}^{\infty}<0$. The blue shaded areas from light to dark correspond to $\epsilon_{+}^{\infty}>0$ for the plasma magnetization parameter $\sigma_0=1, 5, 10,100$. The thick black solid line, red dashed line, and dot-dashed line correspond to the radius of the outer horizon, photosphere, and the outer infinite redshift surface, respectively. We set $M=1$.}\label{Fig31}
\end{center}
\end{figure}
To study energy extraction via the magnetic reconnection mechanism, it is essential to clarify the possible parameter regions where energy extraction can occur. We have combined the outer horizon, photosphere, outer infinite redshift surface, and the energy conditions from Eq.(\ref{condition}) to create Fig.\ref{Fig3} and Fig.\ref{Fig31}, further analyzing the effects of the plasma magnetization parameter $\sigma_0$, the rotating parameter $a$ and deformation parameter $\eta$ on these possible parameter regions of the phase space $(a,r)$ and $(\eta,r)$.
The blue shaded areas from light to dark in the figures represent the regions corresponding to the plasma magnetization parameter $\sigma_0=1, 5, 10,100$ that satisfy the condition $\epsilon_{-}^{\infty}<0$. Fig.\ref{Fig3} and Fig.\ref{Fig31} shows the possible regions expand with increasing parameter $\sigma_0$ which is the same as in the Kerr case. A portion of the blue shaded areas lies below the outer horizon, indicating that their radii are smaller than the outer horizon's. Therefore, this part cannot achieve energy extraction and should be excluded from the possible regions. We primarily focus on the blue shaded areas with radii situated in the ergosphere, between the outer horizon and the infinite redshift surface.
In Fig.\ref{Fig3}, for the case with $\eta\leq0$, under the condition of the ergosphere, most of the blue shaded areas on the right and below that are located without the outer horizon must be excluded. As the deformation parameter $\eta$ decreases, the possible regions of the phase space $(a,r)$ gradually shift to the upper left and contract, becoming smaller than the possible region in the Kerr case. In this scenario, the dominant X-point location needs to satisfy $r \geq M$ and $a\leq M$.
For the case with $\eta>0$, the ergosphere exists for $a>M$and the possible region expands to $r<M$ and $a>M$, which extends beyond the phase space $(a, r)$ in the Kerr black hole. Especially when $0<\eta<\frac{8}{27}$, as the rotating parameter $a$ increase, the outer horizon and photosphere radii will experience a sharp drop to very small values, and the possible region of the phase space $(a,r)$ will rapidly expand to the range with small $r$. This situation of thick ergosphere and small X-points will yield unique results for energy extraction research in the Konoplya-Zhidenko rotating non-Kerr black hole.
In Fig.\ref{Fig31}, as the rotating parameter $a$ increases, the possible regions of the phase space $(\eta,r)$ gradually shift to the upper right and expand. For a highly rotating black hole with $a>M$ and the positive deformation parameter approaches to zero, the possible region $(\eta,r)$ can expand to a radius $r$ that is very small or even close to zero. Especially when $M<a<\frac{2\sqrt{3}}{3}M$, the radius of outer horizon and photosphere increases abruptly with the deformation parameter, and the possible region $(\eta, r)$ forms a narrow strip near $\eta = 0$.
To conclude, the existence of the deformation parameter $\eta$ in the Konoplya-Zhidenko rotating non-Kerr black hole significantly expands the possible regions for energy extraction via the magnetic reconnection mechanism, extending into the areas where $a>M$ and $r<M$ that have not been involved in the Kerr black hole case.
The energy extraction power and efficiency through magnetic reconnection in this region are likely to exhibit fascinating characteristics.

\section{Power and efficiency via magnetic reconnection}
In this section, we will investigate energy extraction from a Konoplya-Zhidenko rotating non-Kerr black hole via magnetic reconnection and analyze the impact of the deformation parameter $\eta$ on both the energy extraction power and efficiency. It is important to emphasize that both energy efficiency and power depend on the negative energy of decelerated plasma absorbed by a black hole within a unit of time. The power $P_{\mathrm{extr}}$ of energy extraction per enthalpy via magnetic reconnection from Konoplya-Zhidenko rotating non-Kerr black hole can be evaluated as \cite{CA}
\begin{equation}
P_{\mathrm{extr}}=-\epsilon_{-}^{\infty} w_0 A_{\mathrm{in}} U_{\mathrm{in}},\label{power}
\end{equation}
where the reconnection in flow four-velocity $U_{\mathrm{in}}=\mathcal{O}\left(10^{-1}\right)$ and $\mathcal{O}\left(10^{-2}\right)$ for the collisionless and collisional regimes. The cross-sectional area of the inflowing plasma $A_{\text{in}}$ can be estimated as $A_{\text{in}}\sim (r_{\infty}^{2}-r_{ph}^{2})$.
\begin{figure}[ht]
\begin{center}
\subfigure[]{\includegraphics[width=6cm]{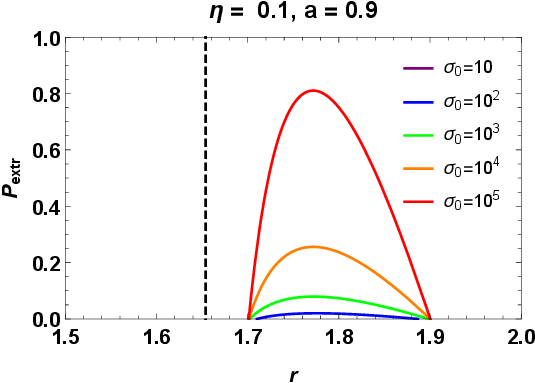}}\subfigure[]{\includegraphics[width=6cm]{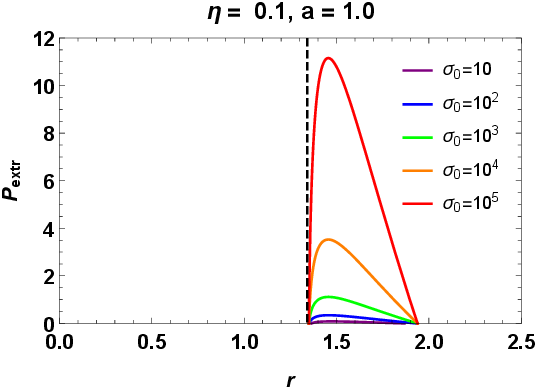}}\\
\subfigure[]{\includegraphics[width=6.2cm]{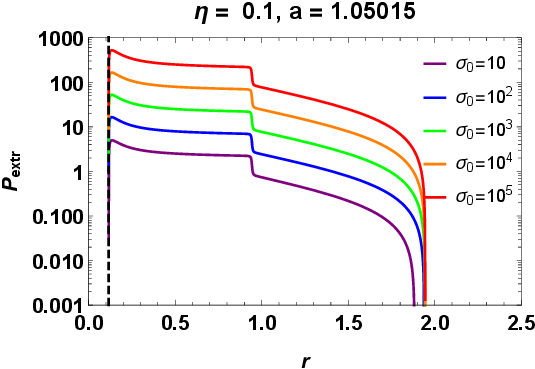}}\subfigure[]{\includegraphics[width=6.2cm]{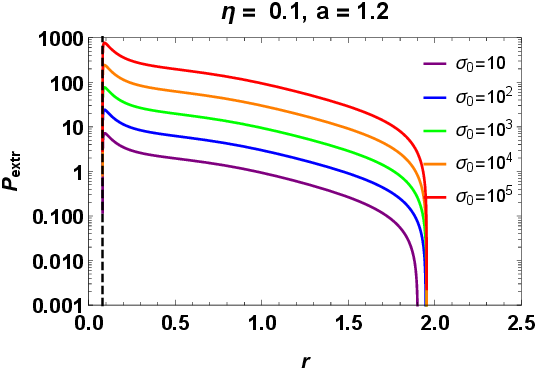}}
\caption{ The change of power $P_{\mathrm{extr}}$ with the dominant X-point location $r$ for different rotation parameter $a$ and plasma magnetization parameter $\sigma_0$. The black dashed lines represent the photosphere $r_{ph}$. We set $M=1$.}\label{Fig4}
\end{center}
\end{figure}
From Eq.(\ref{power}), it can be seen that power is related to four parameters: the deformation parameter $\eta$, the rotation parameter $a$, the dominant X-point location $r$ and the plasma magnetization parameter $\sigma_0$. In Fig.\ref{Fig4}, with the deformation parameter $\eta=0.1$ fixed, the power increases with the plasma magnetization parameter $\sigma_0$ and first increases then decreases with the dominant X-point location $r$. Using the definition of $\eta_c$ mentioned earlier, the critical value of the rotation parameter $a_c$ is determined by solving the equation  $\eta= -\frac{2}{27}\left(\sqrt{4M^2-3a_{c}^2}+2M\right)^2\left(\sqrt{4M^2-3a_{c}^2}-M\right)$. For $a>a_c$, the maximum power approaches a thousand, which is far greater than the tens shown in the Kerr case. This characteristic indicates that the spacetime modifications caused by the deformation parameter $\eta$ have increased the maximum power and are likely to further affect the energy extraction efficiency.
Significantly, in Fig.\ref{Fig4}(c), when $a$ is slightly greater than the critical value $a_c$, the power curve suddenly decreases because the metric component $g_{rr}$ first surges to a large value and then decreases as $r$ increases in this region, causing severe curvature in the radial space. A similar phenomenon occurs in Fig.\ref{Fig2}(d), where there is a sudden change in the energy at infinity per enthalpy $\epsilon_{-}^{\infty}$.

\begin{figure}[t]
\begin{center}
\includegraphics[width=6cm]{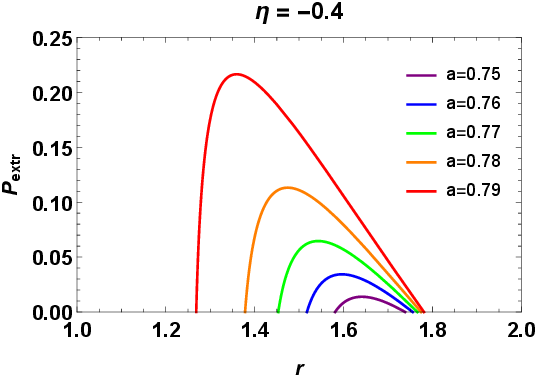}\includegraphics[width=6cm]{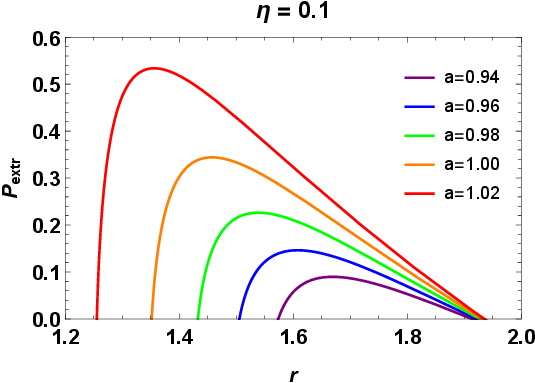}\\
\includegraphics[width=6cm]{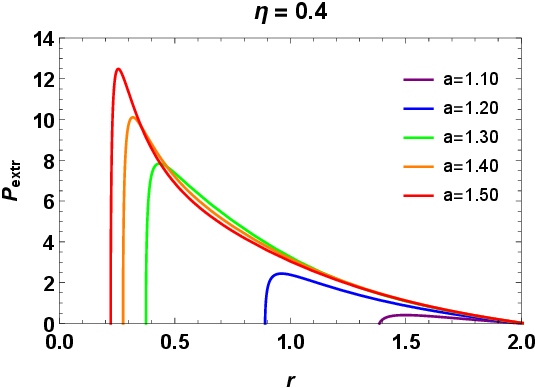}\includegraphics[width=6cm]{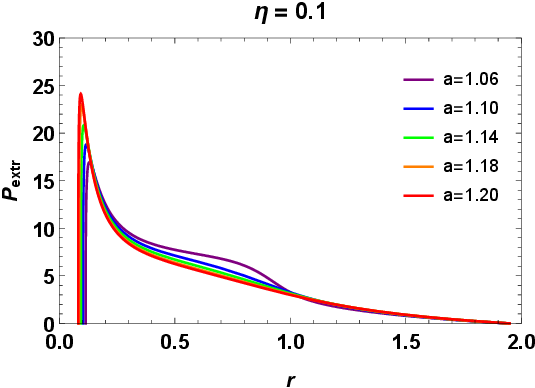}\\
\caption{ The change of power $P_{\mathrm{extr}}$ with the dominant X-point location $r$ for different rotation parameter $a$ and deformation parameter $\eta$. We set $\sigma_0=100$ and $M=1$. }\label{Fig41}
\end{center}
\end{figure}
\begin{figure}[t]
\begin{center}
\includegraphics[width=6cm]{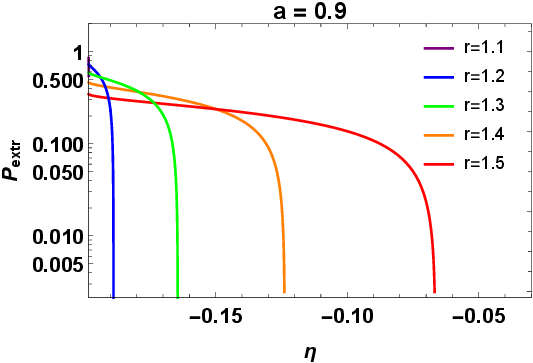}\includegraphics[width=6cm]{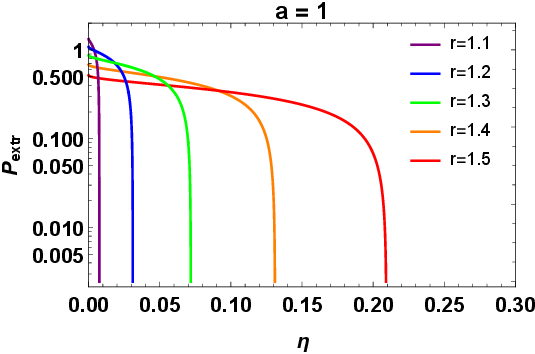}\\
\includegraphics[width=6cm]{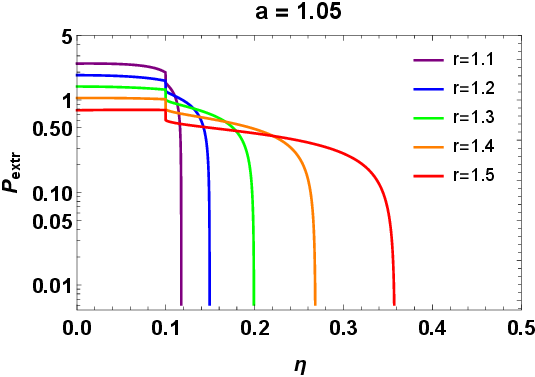}\includegraphics[width=6cm]{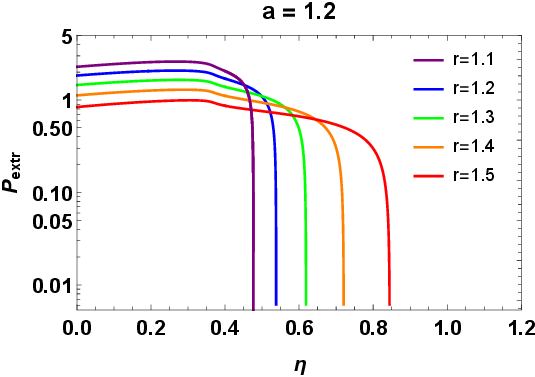}\\
\caption{ The change of power $P_{\mathrm{extr}}$ with the deformation parameter $\eta$ for different rotation parameter $a$ and dominant X-point location $r$. We set $\sigma_0=100$ and $M=1$. }\label{Fig42}
\end{center}
\end{figure}

In Fig.\ref{Fig41}-\ref{Fig42}, with the plasma magnetization parameter $\sigma_0=100$ fixed, we plot the curves of power variation with the dominant X-point location $r$ and the deformation parameter $\eta$. In Fig.\ref{Fig41}, for $\eta<0$, the power increases with the rotation parameter $a$. For $\eta>0$, the peak power significantly increases for large $a$ and continues to rise as $\eta$ increases. In Fig.\ref{Fig42}, the curve drawn from purple to red corresponds to  $r=1.1,1.2,1.3,1.4,1.5$. For the rotation parameter $a\leq M$, the power decreases with the deformation parameter $\eta$. However, for $a>M$, the power initially increases slowly and then sharply decreases to zero with the deformation parameter $\eta$ increases. Notably, when the rotation parameter $M<a<\frac{2\sqrt{3}}{3}M$, the curve exhibits a sudden drop in the middle. This is due to the fact that as the deformation parameter increases, the photosphere radii rapidly grown from a very small value close to $M$, causing a sudden decrease in the cross-sectional area of the inflowing plasma $A_{\text{in}}$ and leading to a sharp drop in the power of magnetic reconnection.

Through the process of magnetic reconnection, most of the magnetic energy is converted into the kinetic energy of the plasma. The plasma with the energy at infinity per enthalpy $\epsilon_{+}^{\infty}$ and $\epsilon_{-}^{\infty}$ escapes from the reconnection layer along the outflow directions of corotating and counter rotating motions.
If the plasma particles with negative energy $\epsilon_{-}^{\infty}$ fall past the outer event horizon into the black hole along the corotating direction while the plasma particles with positive energy $\epsilon_{+}^{\infty}$ escapes to infinity along the counter rotating direction and carry away more energy, the efficiency of energy extraction via magnetic reconnection can be defined as \cite{CA}
\begin{equation}
\chi=\frac{\epsilon_{+}^{\infty}}{\epsilon_{+}^{\infty}+\epsilon_{-}^{\infty}},\label{efficiency}
\end{equation}
\begin{figure}[ht]
\begin{center}
\subfigure[]{\includegraphics[width=6cm]{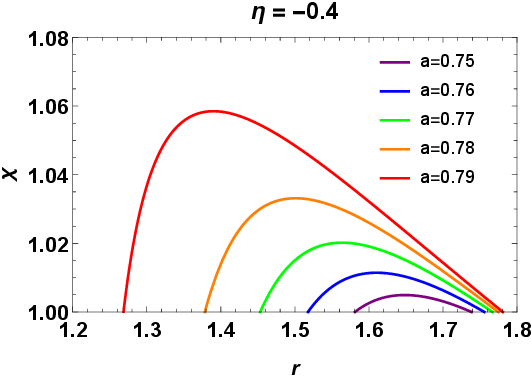}}\subfigure[]{\includegraphics[width=6cm]{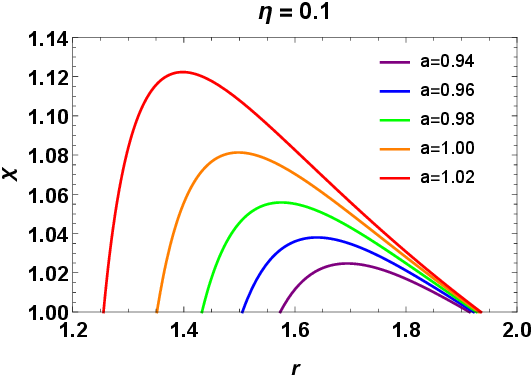}}\\
\subfigure[]{\includegraphics[width=6cm]{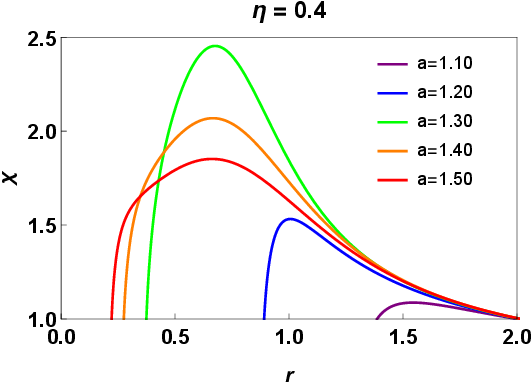}}\subfigure[]{\includegraphics[width=6cm]{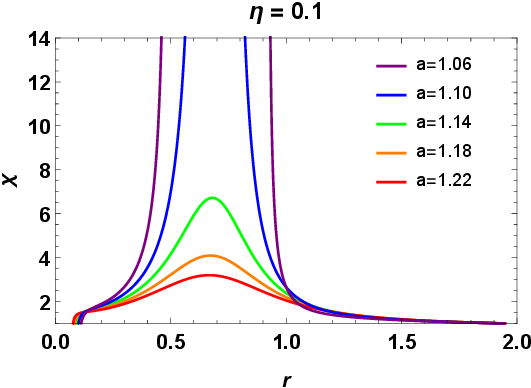}}\\
\caption{ The variation of the efficiency of the energy extraction via magnetic reconnection with the dominant X-point location $r$. We set $\sigma_0=100$ and $M=1$.}\label{Fig5}
\end{center}
\end{figure}

\begin{figure}[ht]
\begin{center}
\subfigure[]{\includegraphics[width=6cm]{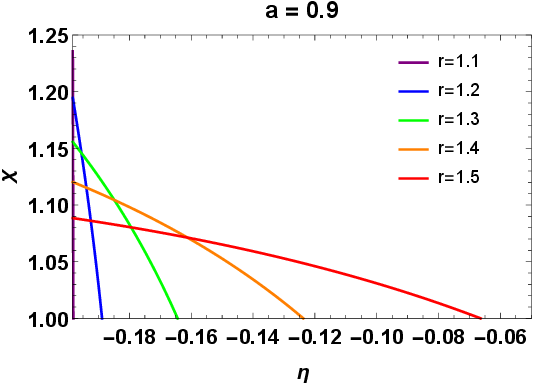}}\subfigure[]{\includegraphics[width=6cm]{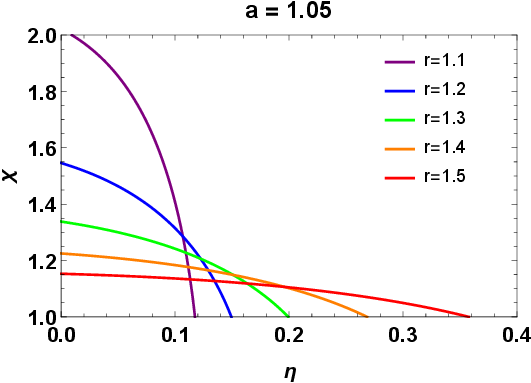}}\\
\subfigure[]{\includegraphics[width=6cm]{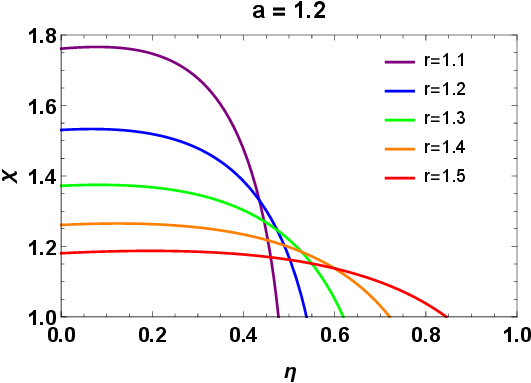}}\subfigure[]{\includegraphics[width=6cm]{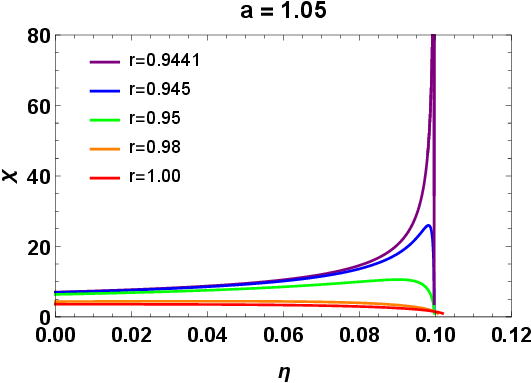}}\\
\caption{ The variation of the efficiency of the energy extraction via magnetic reconnection with the deformation parameter $\eta$. We set $\sigma_0=100$ and $M=1$.}\label{Fig6}
\end{center}
\end{figure}
In Fig.\ref{Fig5}-\ref{Fig6}, we plot the variation of the efficiency of the energy extraction via magnetic reconnection with the dominant X-point location $r$ and the deformation parameter $\eta$ from a Konoplya-Zhidenko rotating non-Kerr black hole. From Fig.\ref{Fig5}(a)-(c), we observe that the efficiency of energy extraction initially increases and then decreases with the dominant X-point location $r$, resulting in a peak in the middle of the curve. When the deformation parameter $\eta \leq 0$, this peak increases with $a$. However, when $\eta > 0$, the rotation parameter $a$ can extend beyond the region where $a > M$, and the peak initially rises then decreases with $a$ .
In Fig.\ref{Fig5}(d), when $0<\eta\leq \frac{8}{27}$ and $a$ is slightly greater than $a_c$, the curve exhibits a distinctive phenomenon of discontinuous intervals, with the endpoints of these intervals showing a sharp increase in efficiency towards infinity. This occurs because $\epsilon_{+}^{\infty} + \epsilon_{-}^{\infty}$ and the efficiency becomes negative within the discontinuous interval, while at the discontinuous endpoints, $\epsilon_{+}^{\infty} + \epsilon_{-}^{\infty}$ tends towards zero and the efficiency  approaching an almost unlimited value. In Fig.\ref{Fig6}(a)-(c), the curve drawn from purple to red corresponds to $r=1.1,1.2,1.3,1.4,1.5$. At the fixed dominant X-point location $r>M$, the efficiency decreases with the deformation parameter $\eta$ in the cases of $a=0.9$ and $a=1.0$, but it first increases and then decreases with the deformation parameter in the cases of $a=1.05$ and $a=1.2$. In Fig.\ref{Fig6}(d), when the rotation parameter $M<a\leq\frac{2\sqrt{3}}{3}M$ and  the dominant X-point location $r$ is slightly greater than $r_c$, the efficiency increases first and then decreases with the deformation parameter, and its peak can reach so high that it is almost unlimited as shown in Fig.\ref{Fig5}(d). The energy extraction efficiency under magnetic reconnection in Kerr black hole can only approach $3/2$\cite{CA}, indicating that the existence of deformation parameter $\eta$ in the Konoplya-Zhidenko rotating non-Kerr black hole greatly raises the upper limit of energy extraction efficiency under magnetic reconnection. The Penrose process in the Konoplya-Zhidenko rotating non-Kerr black hole and the MPP in the parameterized Konoplya-Rezzolla-Zhidenko black hole also yielded similar results, with the maximum efficiency reaching a remarkably high value only if the splitting process occurs very close to the black hole's event horizon\cite{KZ4,Mpp3}. Since the maximum efficiency of energy extraction via magnetic reconnection requires the dominant X-point location to be close to $r_c$ rather than the extreme condition of event horizon radius. Thus, achieving maximum efficiency through the magnetic reconnection process can be easier than the MPP and the Penrose process in a rotating non-Kerr black hole.

Next, we will compare the power extracted from a Konoplya-Zhidenko rotating non-Kerr black hole via magnetic reconnection with the BZ process. The BZ process extracts rotational energy through a magnetic field threading the event horizon of a black hole, and the rate of energy extraction is expressed as:
\begin{equation}
P_{\mathrm{BZ}} \simeq \kappa \Phi_{\mathrm{BH}}^2\left(\Omega_H^2+\zeta_1 \Omega_H^4+\zeta_2 \Omega_H^6\right),
\end{equation}
where $\zeta_1 \approx 1.38$ and $\zeta_2 \approx-9.2$ are the numerical coefficients and $\kappa\approx 0.05$ is a numerical constant related to the magnetic field configuration. The angular velocity at the event horizon of Konoplya-Zhidenko black hole formulated as $\Omega_H=a/(r _{+}^2+a^2)$.
The magnetic flux passing through one hemisphere of the black hole horizon can be expressed as $\Phi_{BH}=\frac{1}{2}\int_{\theta}\int_{\phi}|B^{r}|dA_{\theta \phi}\sim \pi \int_{0}^{\pi}|B^r|\sqrt{-g}d\theta \sim 2\pi (r_{+}^{2}+a^2)B_0 \sin\xi$. Therefore, the power ratio between magnetic reconnection and the BZ process can be written as \cite{1}
\begin{equation}
\frac{P_{\mathrm{extr}}}{P_{\mathrm{BZ}}} \sim \frac{-4\epsilon_{-}^{\infty} A_{\mathrm{in}} U_{\mathrm{in}}}{\pi \kappa \sigma_0 (r _{+}^2+a^2)^2 \sin ^2 \xi\left(\Omega_H^2+\zeta_1 \Omega_H^4+\zeta_2 \Omega_H^6\right)}.
\end{equation}
\begin{figure}[t]
\begin{center}
\subfigure[]{\includegraphics[width=6cm]{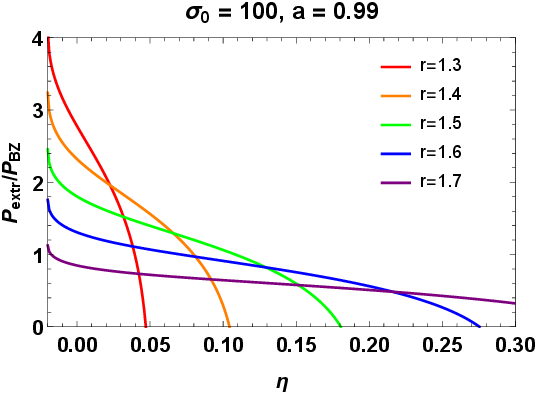}}\subfigure[]{\includegraphics[width=6cm]{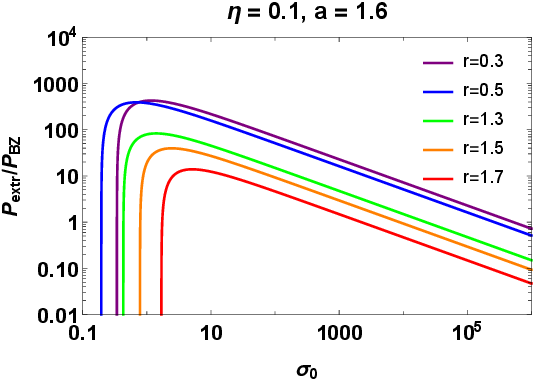}}
\caption{ The variation of the power ratio between magnetic reconnection and the Blandford-Znajek process with the deformation parameter $\eta$ and the plasma magnetization parameter $\sigma_0$. We set $M=1$.}\label{Fig7}
\end{center}
\end{figure}

In Fig.\ref{Fig7}(a),the curve drawn from purple to red corresponds to $r=1.3,1.4,1.5,1.6,1.7$ with $a=0.99$.  For the rotation parameter $a\leq M$, the power ratio between magnetic reconnection and the BZ process decreases with the deformation parameter $\eta$, indicating that the presence of this parameter in the Konoplya-Zhidenko rotating non-Kerr black hole reduces the ratio compared to the Kerr case.
In Fig.\ref{Fig7}(b), we plot the curve of the ratio as a function of the plasma magnetization parameter $\sigma_0$. For the rotation parameter $a>M$ , the dominant X-point location can be extended to small value, and the ratio increases significantly, becoming much greater than that in the Kerr black hole. At the same time, this suggests that magnetic reconnection can be more effective than the BZ process in extracting energy from a superspinning Konoplya-Zhidenko rotating non-Kerr black hole.

\section{Summary}
In this paper we have investigated the energy extraction through the magnetic reconnection from a Konoplya-Zhidenko rotating non-Kerr black hole with an extra deformation parameter, analyzed the influence of deformation parameter on the possible region of the phase space, the power, the efficiency, and the power ratio of energy extraction in magnetic reconnection.

Firstly, to study energy extraction via the magnetic reconnection mechanism, it is essential to clarify the possible parameter regions where energy extraction can occur. For the case with $\eta\leq 0$ , the rotation parameter of the black hole is limited to $a\leq M$, and the possible region of the phase space $(a, r)$ for energy extraction shrinks as the deformation parameter increases. For the case with $\eta>0$, the black hole can exhibit the superspinning case with $a>M$, allowing the phase space to extend into a wider area with $a > M$ and $r < M$, which are beyond the phase space $(a, r)$ in a Kerr black hole. The energy extraction power and efficiency through magnetic reconnection in this expansion region are likely to exhibit fascinating characteristics.

Then we find that the power through the magnetic reconnection mechanism from a Konoplya-Zhidenko rotating non-Kerr black hole increases first and then decreases with the dominant X-point location $r$. For the case with $a\leq M$, the power decreases with the deformation parameter $\eta$. For the superspinning case with $a>M$, the power initially increases slowly and then sharply decreases with $\eta$. The maximum power in the Konoplya-Zhidenko rotating non-Kerr black hole approaches a thousand, which is greater than the Kerr case. Especially when the rotation parameter $M < a < \frac{2 \sqrt{3}  }{3}M$, there is a sudden drop in power due to the sharply increases in the radius of the photosphere as the deformation parameter increases.

The efficiency of the energy extraction via the magnetic reconnection initially increases and then decreases with the dominant X-point location $r$. At the fixed dominant X-point location $r>M$, the efficiency decreases with the deformation parameter $\eta$ in cases with a higher rotation parameter $a$, but in superspinning cases, it first increases and then decreases with $\eta$. When the rotation parameter $M<a\leq\frac{2\sqrt{3}}{3}M$, the maximum efficiency can become nearly unlimited, which is far greater than the maximum efficiency of $3/2$ in Kerr black hole. This indicates that the existence of deformation parameter in a Konoplya-Zhidenko rotating non-Kerr black hole significantly raises the upper limit of energy extraction efficiency. Since the maximum efficiency of energy extraction via magnetic reconnection requires the dominant X-point location close to $r_c$, while the Penrose process requires particle splitting to occur as close as possible to the extremely small event horizon, achieving maximum efficiency through the magnetic reconnection process can be easier than through the MPP and the Penrose process in a rotating non-Kerr black hole.

Finally, we calculated the power ratio between magnetic reconnection and the BZ process. For the case with $a\leq M$, the power ratio decreases with the deformation parameter $\eta$, indicating that the presence of the deformation parameter reduces the ratio in the Konoplya-Zhidenko rotating non-Kerr black hole.
For the superspinning case with $a>M$, the dominant X-point location can be extended to a smaller value and the ratio increases significantly, becoming much greater than the Kerr case. At the same time, this suggests that magnetic reconnection can be more effective than the BZ process in extracting energy from a superspinning Konoplya-Zhidenko rotating non-Kerr black hole.

In conclusion, the positive deformation parameters in the Konoplya-Zhidenko rotating non-Kerr black hole expand the possible region of energy extraction via magnetic reconnection to $a>M$ and $r<M$, improving the maximum power, maximum efficiency, and the maximum ratio between magnetic reconnection and the BZ process of energy extraction. Therefore, the Konoplya-Zhidenko rotating non-Kerr black hole can extract more energy via magnetic reconnection than Kerr black hole. These effects of the deformation parameter may provide valuable clues for future astronomical observations of black holes and verification of gravity theories.

\section{\bf Acknowledgments}
This work was partially supported by the National Natural
Science Foundation of China (Grant Nos. 12205140, 12275078, 11875026, 12035005, and 2020YFC2201400)
and the Natural Science Foundation of Hunan Province (Grant No. 2023JJ40523).

\end{document}